\documentstyle[times,pramana,epsf,floats,graphicx]{ias}
\begin{document}
\mark{{Lattice QCD at nonzero $T$ and $\mu$}{R. V. Gavai}}
\title{Present Status of Lattice QCD at nonzero $T$ and $\mu$
\vskip-1.8cm\hfill\small hep-ph/0607050, TIFR/TH/06-18\vskip1.5cm
}

\author{Rajiv\ V.\ Gavai}
\address{Department of Theoretical Physics, Tata Institute of
Fundamental Research, Mumbai 400 005.} 
\abstract{
I review a few selected topics in Lattice Quantum Chromo Dynamics, focusing
more on the recent results.  These include i) the equation of state and speed
of sound, ii) $J/\psi$ suppression, iii) flavour correlations and iv) the QCD
phase diagram in the $\mu$-$T$ plane. 
}

\maketitle
\section{Introduction}

In the continuous quest for Quark-Gluon Plasma (QGP), which began about two 
decades ago at the SPS at CERN, Geneva, and has continued in a spectacular
way at the Relativistic Heavy Ion Collider (RHIC) at BNL, New York, USA,
one now has a plethora of theoretical suggestions for signals of QGP as
well as huge amounts of data.   The Large Hadron Collider (LHC) at CERN
will take us to even higher colliding energies of heavy ions than before.
In the complex task of putting the pieces of the jigsaw puzzle together to
establish eventually this new phase of strongly interacting matter from the
experimental data, one needs as much of theoretical help as can be
provided.  In particular, theory needs to provide reliable estimates of 
various quantities, such as, the transition temperature, the equation of
state (EoS) and the critical energy density needed to reach the QGP phase.
Information on various properties of QGP, such as, the nature of its 
excitations and strength of their interactions would also be crucial in the 
experimental search.

Quantum Chromo Dynamics (QCD) defined on a space-time lattice, lattice QCD, is
the only successful and reliable tool to extract the desired non-perturbative
physics from the underlying theory.  This first-principles based and
(essentially) parameter-free approach should be contrasted with  the results
from other approaches, primarily models.  Thus not only does lattice QCD lead
us to the phenomenon of quark confinement and spontaneous breaking of chiral
symmetry (or the answer to why pion is so light) but it also provides us with a
quantitative understanding of the spectrum of hadrons and their other
properties. Indeed, these results have been accorded an `iconic' status in
theoretical high energy physics recently \cite{wil}.  One hopes that similar
reliable information on the transition to QGP and properties of QGP will be
provided by these techniques, and any experimental demonstration of a failure
of a prediction of lattice QCD, such as the transition to quark-gluon plasma,
will be tantamount to one of the best experimental evidence for physics beyond
the standard model, discussed rather extensively at this workshop.

Lattice formulation of QCD associates quark fields, $\psi (x)$, and the
antiquark fields $\bar\psi (x)$ with the site $x = (x_1, x_2, x_3, x_4)$ of a
4-dimensional hypercubic lattice.  The (inverse) lattice spacing $a$ acts as
the ultra-violet cut-off.  Continuum limit of vanishing $a$ corresponds to
removal of the cut-off.  As in the case of the continuum field theory, one
obtains a lattice gauge theory by demanding invariance of the Lagrangian for
free quark-antiquarks (e.g. obtained by a straightforward discretization of the
usual Dirac Lagrangian) under any {\it local} phase rotation of these fields.
This can be accomplished by introducing lattice gauge fields $U^\mu_x \equiv
U_\mu (x)$ which are associated with a directed link from the site $x$ to $x +
\hat\mu a$. 

Defining a partition function ${\cal Z}$ for these fields, akin to a 
complicated version of the familiar Ising model partition function, 
\begin{eqnarray}
{\cal Z} &=& \int \prod_{x,\hat\mu}\ dU_\mu (x) \prod_x\ d\psi(x)\ d\bar\psi
(x) e^{-S_G-S_F} \nonumber \\ [2mm]
&=& \int \prod_{x,\hat\mu}\ \prod_f \det~M (am^{sea}_f,a \mu_f) e^{-S_G} , 
\label{eq:four}
\end{eqnarray}  
\noindent 
where $S_G$ and $S_F$ are gluonic and quark actions, $M$ is the Dirac matrix in
$x$, color, spin, flavour space for fermions of mass $am^{sea}_f$ and $a\mu_f$
is the chemical potential (in lattice units).  One can compute quantum
expectation values of any physical observable $\Theta$, which may contain
fermion propagators of mass $am^{valence}$, as averages with respect to the
${\cal Z}$ above.  Thus, e.g., masses of physical particles are obtained from
the exponential decays of appropriate correlation functions. Of relevance to
heavy ion physics, is the investigation of eq. \ref{eq:four} at finite
temperature, which is achieved by taking a $N_s^3 \times N_t$ lattice, where
$N_s$ is the number of lattice sites in a space direction and $N_t$ in the time
direction.  This leads to volume $V=N_s^3.a^3$ and temperature $T=(a N_t)^{-1}
$.  Clearly, one needs to have $ N_s \gg N_t$.

The Monte Carlo technique to evaluate  the expectation value of any physical
observable, consists of the following 3 steps : 1) Generate as large an
ensemble of sets of links $\{U^\mu_x\}$ for the whole lattice as possible, such
that each set of $\{U^\mu_x\}$ occurs with a probability proportional to
$\prod_f \det M_f \cdot \exp [- S_G (\{U^\mu_x\})]$, 2) Evaluate the observable
for each configuration $\{U^\mu_x\}$ and 3) Take its average over all the
configurations in the set.  Due to the enormity of the computational task to
generate the set of $\{U^\mu_x\}$ for full QCD, i.e., for a theory with all
virtual quark loops included, one employs increasingly severe approximations
with decreasing amount of computer time.  These are (i) full QCD simulations on
smaller lattices, (ii) partially quenched QCD simulations with $am^{sea}$ large
and greater than $am^{valence}$ and (iii) quenched QCD simulations with
$am^{sea} = \infty$ (i.e. no dynamical quarks).  The early lattice results and
today's best results are obtained in the quenched approximation. However, some
aspects, such as the order of the phase transition, do depend strongly on the
dynamical quark content, necessitating a judicious use of the quenched
approximation.

A transition temperature of $T_c \sim 170$ MeV for two flavour QCD has been
earlier estimated \cite{cppax,bielefeld}.  The transition seems to be
continuous.  For 3 light flavours,the transition temperature seems to be lower
by about 20 MeV \cite{bielefeld},  bracketing our world of two light and one
heavy flavours, by that amount.  Similarly, equation of state (EOS) has been
predicted by lattice QCD.  Other quantities, notably the
Wr\'oblewski Parameter $\lambda_s$, which measures the  strangeness enhancement
in heavy ion physics, have also been predicted \cite{gg02} by lattice QCD.
These earlier results were obtained either in quenched approximation or on
coarser $N_t$ =4 lattices for full theory.  Thrust of the new results in the
recent past has been on i) employing larger lattices  to achieve continuum
limit and to lighter quarks, ii) more complex observables, such as speed of
sound, transport coefficients, fluctuations and susceptibilities,
$J/\psi$-dissolution/persistence, dileptons etc. and iii) theoretically more
challenging $T$-$\mu$ phase diagram.  An interesting comparison of 
predictions of conformally invariant QCD-like theories with lattice results
has also been made.  

For reasons of both time and interest, I have chosen to limit this review to a
few of the above mentioned selected topics.  A quick overview of the basic
lattice gauge theory can be found in \cite{qcd02} or many textbooks.  In the
next sections, I intend to discuss only the recent results, leaving out many
technical details.  A short summary is provided at the end.  Let me emphasise
here that one has witnessed recently a lot of activity in model building to
explain the lattice QCD results.  These include Quasiparticle models, Hadron
Resonance Gas, Quarkonia from Lattice $Q \bar Q$ potential, sQGP and coloured
states.  Many of them can, and eventually did, form interesting topics for
working group discussions, leading even to a paper on the archive eventually.

\section{Properties of QGP}

In this section I shall review the recent progress made in the past few years
in pinning down various properties of QGP using the lattice approach.  These
have been chosen for their direct connection with the experimental heavy ion
physics.

\subsection{Equation of State and Speed of Sound}

\begin{figure}[htb]
\begin{center}
\begin{minipage}{0.48\textwidth}
\epsfxsize=6cm
\centerline{\epsfbox{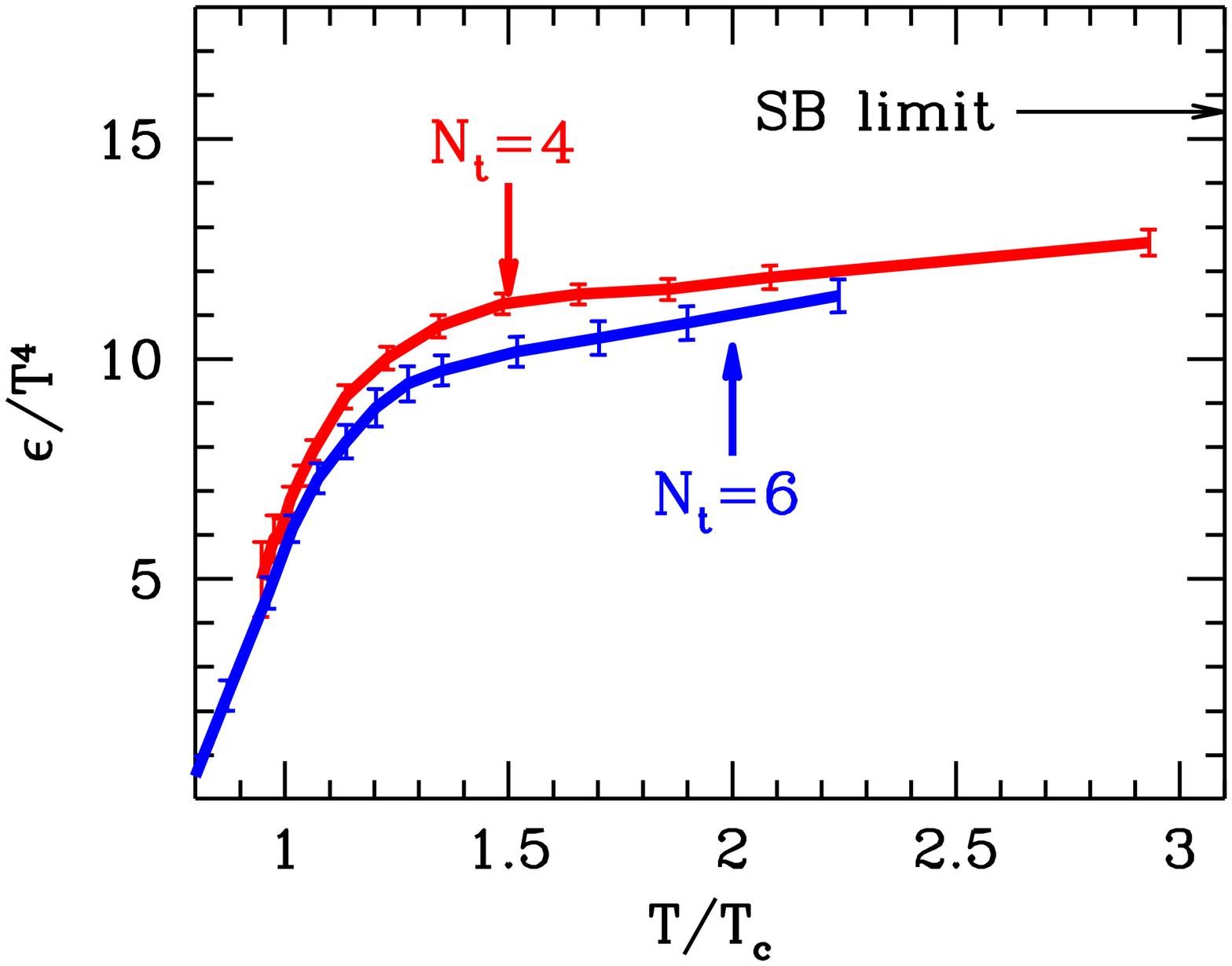}}
\end{minipage}\hfill
\begin{minipage}{0.48\textwidth}\raggedright
\epsfxsize=6cm
\centerline{\epsfbox{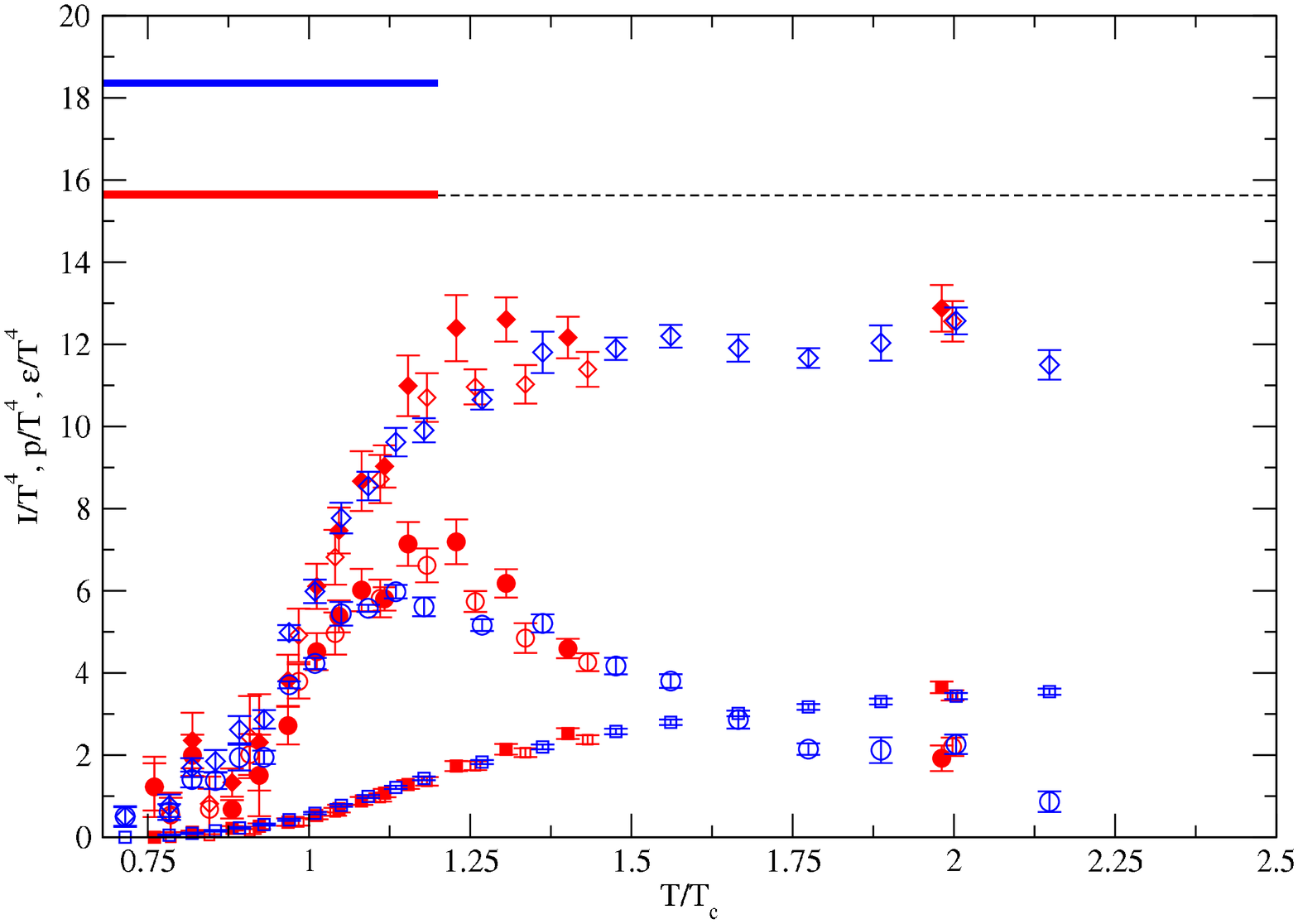}}
\end{minipage}
\caption{{Recent results for EoS on $N_t$ = 6 lattices from Refs. [6] (left) 
and [7] (right).}}
\label{fg:one}
\end{center}
\end{figure}

Equation of state plays a critical role in a variety of aspects of heavy ion
collisions.  Thus estimates of the energy density at $T_c$ are required to
choose the appropriate heavy ion colliding energy.  Recently results for the
pressure and energy density for $N_t=6$ lattices at nonzero temperature QCD
with 2+1 flavors of improved staggered quarks were reported \cite{milc05}.
Using improved gauge action and improved staggered quark action to eliminate
the cut-off effects at ${\cal O}(a^2)$ and with the heavy quark mass $m_s$
fixed at approximately the physical strange quark mass but with the two
degenerate light quark masses $m_{ud} =0.1m_s$ or $0.2m_s$, it was found that
the results were in agreement with those on $N_t =4 $ lattices, as seen on the
right panel of Figure \ref{fg:one}. 

Using another improved action, similar results were obtained by another group
\cite{wup05} for quark masses corresponding to physical meson masses.  These
are displayed in the left panel of Figure \ref{fg:one}.  Again, one sees small
changes compared to the $N_t$ = 4 case. In particular, the critical energy
density, $\epsilon(T_c) \sim 6 T_c^4$ still.  One note of caution though is
that the spatial volumes in physical units (such as inverse pion mass) are
rather small and some changes may therefore be expected in true thermodynamic
limit.

Velocity of sound, $C_s$,  in QGP and dense hadronic media is a crucial input
to many phenomenological studies based on the hydrodynamical approach. The
presence of elliptic flow was established in this way and has been regarded as
a key evidence in favour of the collective behaviour of the produced matter.
Estimates of $C_s$ in the transition region have recently been obtained
using a new approach \cite{swag} which relates it to the temperature derivative
of the anomaly measure $\Delta/\epsilon$, where $\Delta = \epsilon - 3P$.
Combining further with a new method to obtain $\epsilon$ and $P$ by an improved
operator method, which leads to positive pressure on all lattices at all
temperatures, $C_s$ as well as the specific heat $C_v$ were obtained
\cite{swag2} in the continuum limit.

\begin{figure}[htb]
\begin{center}
\begin{minipage}{0.48\textwidth}
\epsfxsize=6cm
\centerline{\epsfbox{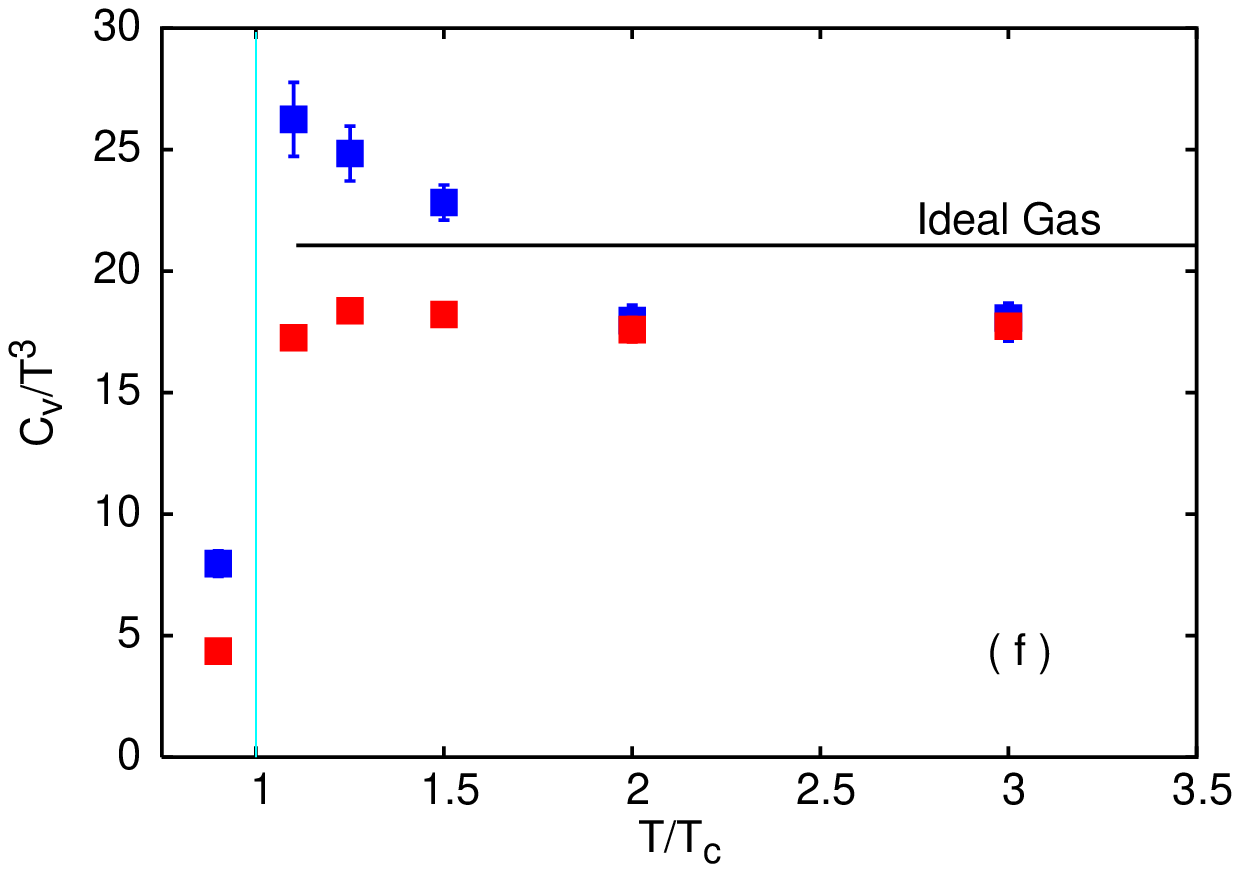}}
\end{minipage}\hfill
\begin{minipage}{0.48\textwidth}\raggedright
\epsfxsize=6cm
\centerline{\epsfbox{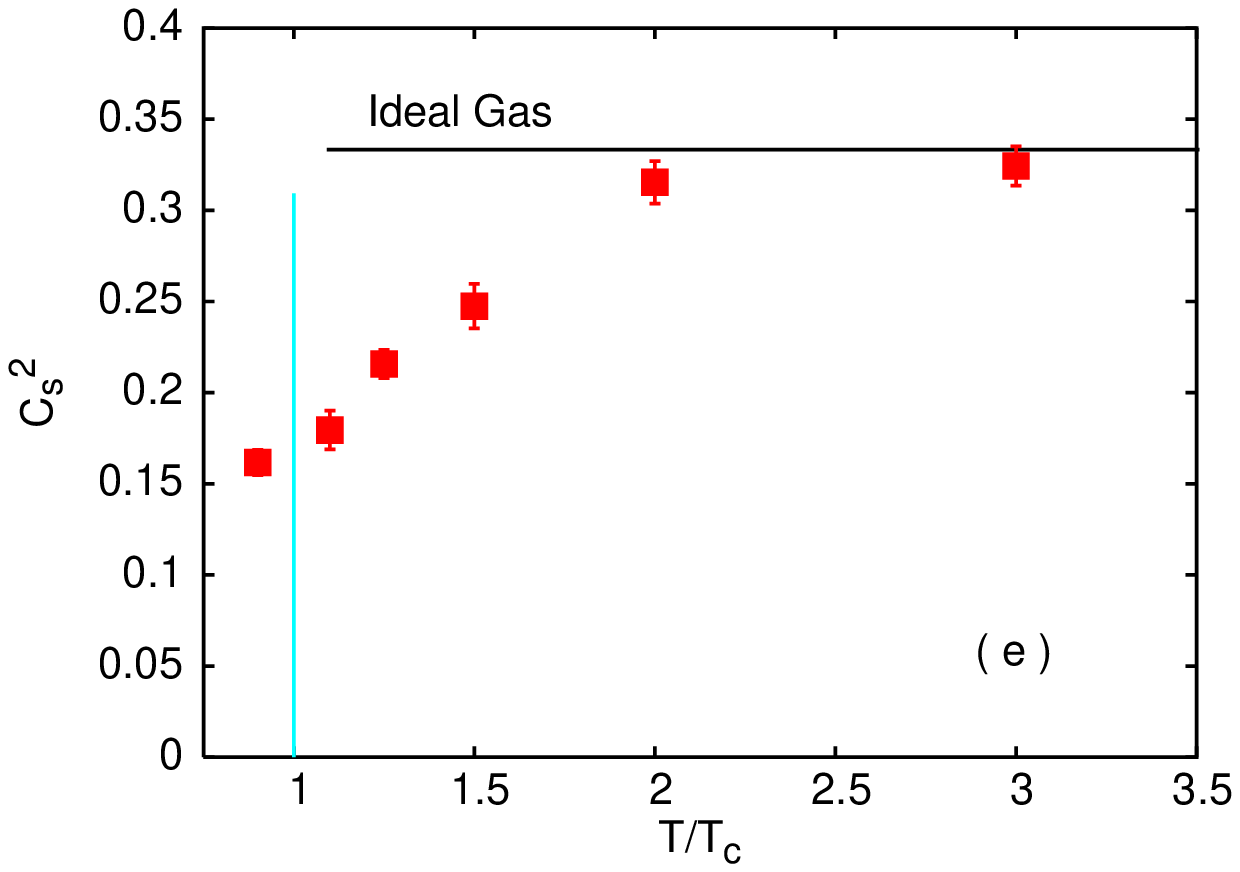}}
\end{minipage}
\caption{ $C_v/T^3$ and $4 \epsilon/T^4$ (left) and $C_s^2$ (right)
as a function of $T/T_c$ for quenched QCD in continuum from Ref. [9].}
\label{fg:two}
\end{center}
\end{figure}

Lattices with large temporal extent, $N_t$ =8,10 and 12, and spatial sizes up to
$N_s$= 38 were used in the quenched approximation to obtain the results shown
in Figure \ref{fg:two}. One sees in the left panel that $C_v \sim 4 \epsilon$
for $T \ge 2T_c$ but its value is not close to the ideal gas limit.  On the
other hand, the speed of sound shown on the right, $C_s^2$, is close to the
ideal gas limit by 2$T_c$. Interestingly, it does not seem to show any
structure near $T_c$, whereas the specific heat does does hint at a peak at
$T_c$.  It has been argued that fluctuations in $p_T$ may be able to unravel
the specific heat and its peak from heavy ion data.  More theoretical studies
are needed to sharpen this idea and to obtain a precise experimental measure.
The above lattice results are encouraging for such studies.

An exciting and curious agreement of the  entropy density $s$ computed for
quenched continuum QCD has been reported \cite{swag2} recently with that
obtained using strong coupling prediction in the supersymmentric Yang-Mills
theory in the temperature range of 2-3 $T_c$.  The latter predicts \cite{kleb}
$s/s_0 = f(g^2 N_c)$, where $f(x) =
\frac34+\frac{45}{32}\zeta(3)x^{-3/2}+\cdots $ and $s_0 = \frac23 \pi^2
N_c^2T^3$ is the ideal gas entropy density.  In contrast, $s_0 = \frac{4}{45}
\pi^2 (N_c^2-1) T^3$ for quenched QCD (i.e, without supersymmetry).
Nevertheless, the prediction does seem to do well both in normalization and the
shape in spite of the badly broken supersymmetry for the lattice results.
Understanding this and extending it to even lower temperatures where weak
coupling methods do fail would be very exciting indeed.

\subsection{ QGP -- (Almost) Perfect Liquid ?}

\begin{figure}
\begin{center}
\epsfxsize=8cm
\centerline{\epsfbox{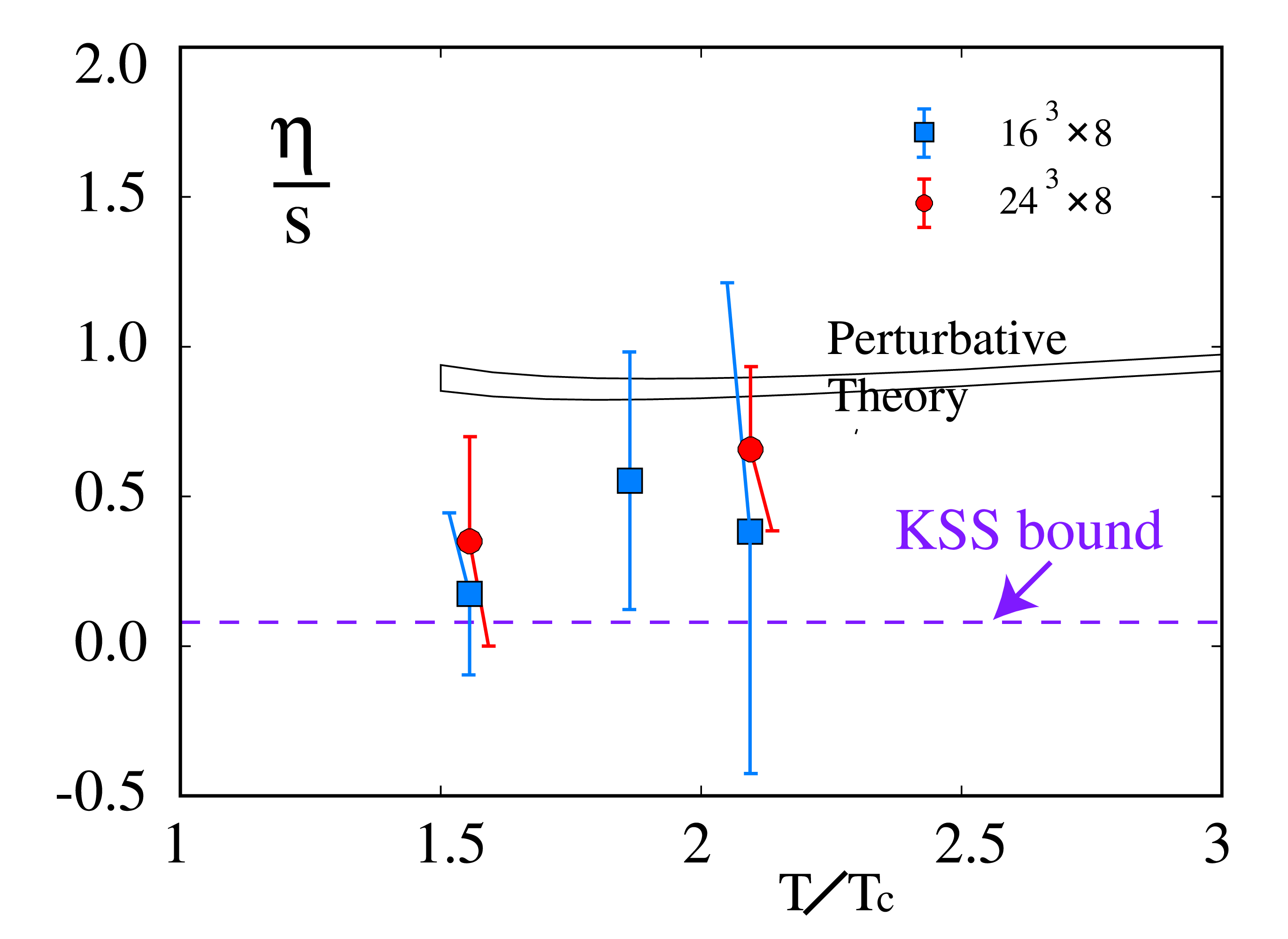}}
\caption{Ratio of shear viscosity and entropy density for quenched 
QCD [13].}
\label{fg:three}
\end{center}

\end{figure}

From the measured azimuthal angle distribution of various particles produced in
a collision, one extracts a quantity called elliptic flow, $v_2$.
Phenomenological studies of $v_2(p_T)$ show it to be consistent with QGP
displaying ideal hydrodynamical behaviour \cite {jyo}.  Indeed, from the small
deviations, one can derive \cite{teany} a bound on the shear viscosity, $\eta$,
of QGP: $\Gamma_s/\tau_0 = \frac43 \eta/s T \tau_0 \le 0.1$, where $\tau_0$ is
the formation time.  This suggests a very small value for the dimensionless
ratio, $\eta/s$, whereas perturbation theory leads to a large value for it,
giving rise to the description of QGP produced at RHIC as being an almost
perfect and strongly coupled liquid.

Kubo's linear response theory permits the determination of the transport
coefficients, such as $\eta$, in terms of equilibrium correlation functions.
In particular, one needs the correlation function of the energy-momentum
tensor.  This is obtained on lattices in $x$-space and transformed to the
momentum space, i.e, obtained at discrete Matsubara frequencies.  In order to
extract the transport coefficients, which are determined by the behaviour of
the retarded correlation functions at small frequencies $\omega$, one continues
these in the complex $\omega$-plane.   Figure \ref{fg:three} displays the
latest results \cite{ns} on the desired ratio $\eta/s$ on lattices as large as
$24^3 \times 8$ but in QCD without dynamical quarks.   Also shown are the
estimates from perturbation theory and an analytic bound from supersymmetric
QCD which is fairly close to the bound from hydrodynamics quoted above. 
The lattice results start from being close to these bounds in the vicinity
of the transition and show a tendency of going towards the larger perturbative
value as temperature increases.  Calculations on larger lattices and with
inclusion of dynamical quarks are needed to confirm these results and to make
them more precise.   Nevertheless their proximity to the bound from the
heavy-ion data is very encouraging.

\subsection{ Anomalous $J/\psi$ Suppression.}

\begin{figure}[htb]
\begin{center}
\begin{minipage}{0.49\textwidth}
\epsfxsize=6.0cm
\centerline{\epsfbox{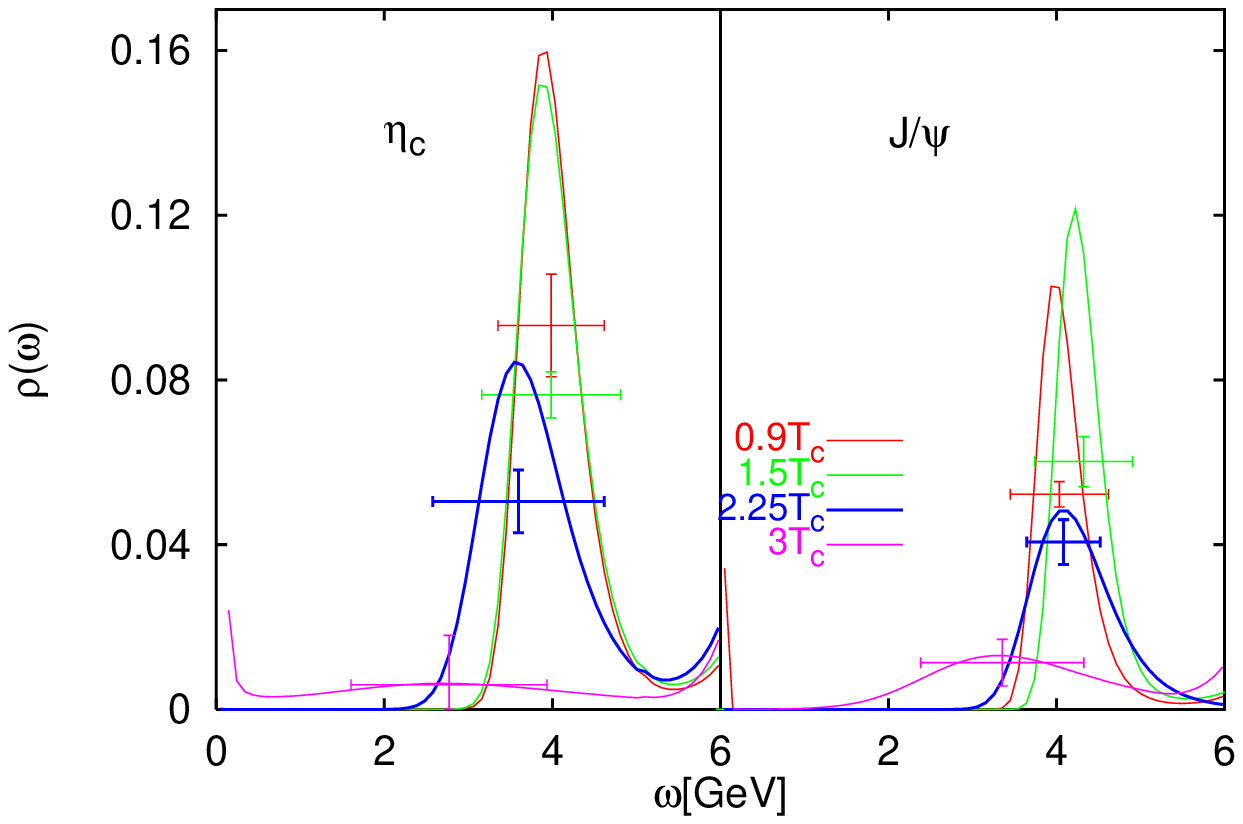}}
\end{minipage}\hfill
\begin{minipage}{0.49\textwidth}\raggedright
\epsfxsize=6.0cm
\centerline{\epsfbox{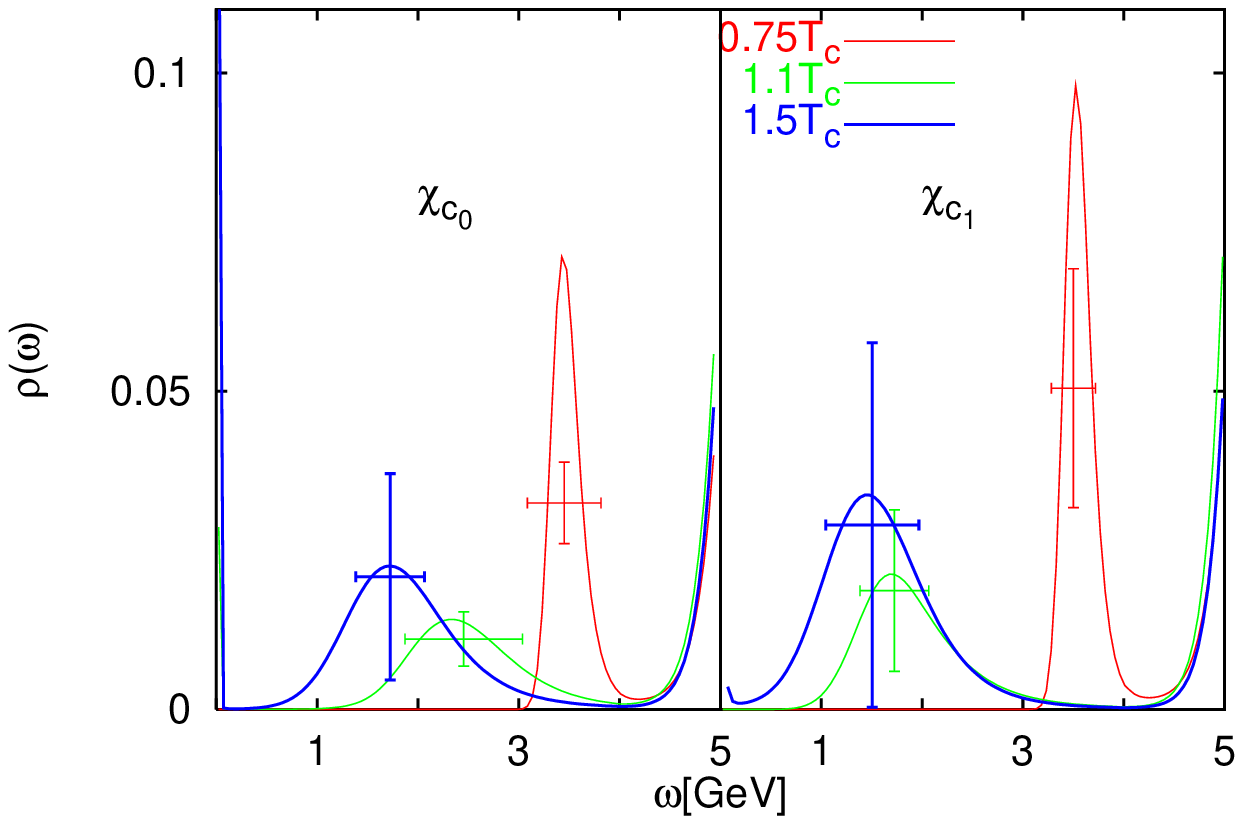}}
\end{minipage} 
\end{center}
\caption{Spectral functions of $\eta_c$ , $J/\psi$(left) and 
the $\chi$ states (right) in quenched QCD [16] at various 
temperatures indicated. }
\label{fg:four}
\end{figure} 

Suppression of the most famous quarkonium, $J/\psi$, has been widely known as a
signal for QGP production in heavy-ion collisions and a lot has been learnt
\cite{wh06} since the early data on this subject.   Recently, the NA50
collaboration from CERN came out with their precise results. Moreover,
$J/\psi$-suppression has been observed at the RHIC in BNL as well.  To briefly
summerise these latest results \cite{jpsi} from CERN and BNL and their
comparison with model predictions,
\begin{itemize}

\item Sulphur-Uranium and peripheral lead-lead (Pb-Pb) results for the ratio of
$J/\psi$ production cross section to that of Drell-Yan behave as expected from
extrapolations of proton-nucleus data.  Glauber model with an absorption cross
section of 4.18 $\pm$ 0.35 mb for the $J/\psi$, determined from p-A data,
decided these expectations.

\item Pb-Pb central data on the other hand show anomalous suppression with
respect to the corresponding expectations.

\item The $\psi'/$DY ratio is {\bf not} compatible for even S-U and
peripheral Pb-Pb with similar expectations from proton-nucleus (with an
absorption cross section of 7.6 $\pm$ 1.12 mb).

\item $\psi'$ anomalous suppression sets in thus earlier than $J/\psi$.

\item theoretical models that were successful in describing the above SPS,
CERN data predicted too much suppression at RHIC compared to what was
observed by PHENIX at RHIC.

\end{itemize}

The impressive sets of data from CERN and RHIC have attracted a lot of
attention from lattice experts.  The original Matsui-Satz idea of $J/\psi$
suppression was based on simple quarkonium potential models and an ansatz for
the temperature dependence of the potential.   It seems very important to check
whether similar conclusions follows from the underlying theory, QCD, with as
few assumptions as possible.   Recognition of the so called MEM (Maximum
Entropy Method) as a tool to extract the spectral functions of mesons from the
temporal correlators computed on the lattice has permitted a fresh assessment
of the original idea.   Figure \ref{fg:four} exhibits the spectral functions
\cite{saumen} of the $\eta_c$, $J/\psi$ and the $\chi$ states in quenched QCD
at various temperatures.

As seen on the right panel, the peaks for $\chi_c$ do not show any significance
beyond the error bars by 1.1$T_c$, i.e, they do seem to dissolve in agreement
with the potential model estimate but the left panel shows persistence of
$J/\psi$ and $\eta_c$ up to 2.25 $T_c$; they seem to melt away only by 3$T_c$.
Similar results have been obtained by other groups \cite{ah}, although
differences persist on the precise melting temperature for $J/\psi$.   In
addition to the need to iron them out, requiring larger lattices, uniform
criteria and a comparison at the correlator levels, one has to also include
dynamical quarks in these computations in quenched QCD.  There is a lot more
work to do but one can still ask whether these results should lead to changes
in expectations of the suppression patterns as a function of temperature or the
colliding energy since only a fifth to a third of the  observed $J/\psi$ come
from the states which do melt soon after the transition.  Another interesting
question is about quarkonia moving in the heat bath.   One may expect them to
see more energetic gluons, leading to more dissociation at the same temperature
as the momentum increases.  Preliminary results \cite {saum2} show this to be
indeed true for even $J/\psi$ and $\eta_c$.   However, the effect seems
significant both below and above the transition, leading one to wonder whether
it plays any role in the anomalous suppression seen in the heavy-ion
collisions.

\section{QCD Phase Diagram}

Lattice QCD at nonzero baryon density may help us in understanding, or even
deriving, an interesting physical phenomena, namely, color superconductivity,
which may find applications in the astrophysics of strange quark stars.  From a
theoretical viewpoint, it is, of course, crucial in completing the $\mu_B$-$T$
phase diagram of QCD.  Both the numerical and the analytical methods used at
finite temperature are inadequate in this case due to the fact that the
fermionic determinant det $M(\mu)$ is complex for $\mu \ne 0$, commonly
referred to as the sign (or the phase) problem.  Various models, notably the
Nambu-Jona-Lasinio or the random matrix model has played a big role in shaping
our understanding of the QCD phase diagram.  However, there have been some new
exciting developments in the recent past for small $\mu$.  Most earlier
attempts comprised of exploring first the zero temperature axis, where the
problem is hardest.  Recognizing this the latest strategy has been to work for
small $\mu$ in the vicinity of the quark-hadron transition, and study its
behaviour as $\mu$ is turned on.  Various methods \cite{nzmu,nzmu1}, such as the
re-weighting method, a Taylor expansion in $\mu$, analytic continuation from
imaginary $\mu$, have lead to similar qualitative results.  I will provide a
flavour of these results by briefly mentioning here our results obtained by
Taylor expansion, referring the reader to the original works in 
\cite{nzmu,nzmu1} for a detailed comparison.

The Taylor expansion method has several advantages over the others.  Prime
amongst them is the ease of taking continuum and thermodynamic limit.  As
mentioned in Introduction, one {\em has} to take these limits for the results
to have any relevance to the real world of experiments.  The re-weighting
method, for example, has a factor that grows exponentially with lattice size
and also has no systematic control over discretization errors.  Analytic
continuation to real $\mu$ is also done term by term in a small
$\mu$-expansion. Employing the Taylor expansion, we \cite{endpt} studied volume
dependence at several $T$ to i) bracket the critical region and then to ii)
track its change as a function of volume.  A strong volume dependence was found
which changed the critical point obtained earlier by other methods
substantially.  The lattices we used were  4 $\times N_s^3$, with $N_s=$ 8, 10,
12, 16, 24, enabling us to vary the volume $V = N_s^3 a^3$ at fixed
temperature, i.e, fixed $a$.  Our dynamical simulations for staggered fermions
with two light dynamical ($N_f =2$ of mass $m/T_c =0.1$) quarks were made using
the well-known R-algorithm with a trajectory length scaled $\propto N_s$. From
previous work \cite{milctc}, it is known that the transition temperature is
$T_c/m_\rho = 0.186 \pm 0.006$ and the choice of our quark mass corresponds to
$m_\pi/m_\rho = 0.31 \pm 0.01$.  While still high compared to the real world,
it is one of the smallest pion mass so far used; lowering pion mass further
necessitates even larger volumes than we were able to employ.  Our
simulations were made at $T/T_c = $ 0.75(2), 0.80(2), 0.85(1), 0.90(1),
0.95(1), 0.975(10), 1.00(1), 1.045(1), 1.15(1), 1.25(2), 1.65(6) and 2.15(10).
Typical statistics used was 50-100 in (max) autocorrelation units.

Defining $\mu_f$ as the chemical potential for a flavour $f = u, d, s$ and
$\mu_0 = \mu_u + \mu_d + \mu_s$ and $\mu_3 = \mu_u - \mu_d$ as baryon and
isospin chemical potentials, the respective density and susceptibility can be
obtained from eq. \ref{eq:four} as :  
\begin{equation}
\qquad \qquad n_i =  \frac{T}{V} {{\partial \ln {\cal Z}}\over{\partial \mu_i}}, \qquad
\chi_{ij} =  \frac{T}{V} {{\partial^2 \ln {\cal Z}}\over{\partial \mu_i \partial \mu_j} } 
\label{defs}
\end{equation}

Setting $\mu_f$ = 0 after taking the derivatives, $n_f$ = 0 but 
$\chi_{ij}$ are nontrivial. The diagonal $\chi$'s are found \cite{sus1} to be
\begin{eqnarray}
\chi_0 &=& \frac{1}{2} [ {\cal O}_1(m_u) + \frac{1}{2} {\cal O}_2(m_u) ] \\
\chi_3 &=& \frac{1}{2}  {\cal O}_1(m_u)  \\
\chi_s &=& \frac{1}{4} [ {\cal O}_1(m_s) + \frac{1}{4} {\cal O}_2(m_s) ] 
\label{chidef}
\end{eqnarray}  
Here ${\cal O}_i$ are trace of products of $M^{-1}$, $M'$ and $M''$ and are 
estimated by a stochastic method: 
$ {\rm Tr}~A = \sum^{N_v}_{i=1} R_i^\dag A R_i / 2N_v~~,$ and
$ ({\rm Tr}~A)^2 = 2 \sum^{L}_{i>j=1} ({\rm Tr}~A)_i ({\rm Tr}~A)_j/ 
L(L-1)$,
where $R_i$ is a complex vector from a set of $N_v$, subdivided in L
independent sets. Further details can be found in \cite{endpt,sus1}.

Denoting higher order susceptibilities by  $\chi_{n_u,n_d}$,
the pressure $P$ has the expansion in $\mu$:
\begin{equation}
   \frac{\Delta P}{T^4} \equiv \frac{P(\mu, T)}{T^4} - \frac{P(0, T)}{T^4}
   = \sum_{n_u,n_d} \chi_{n_u,n_d}\;
	\frac{1}{n_u!}\, \left( \frac{\mu_u}{T} \right)^{n_u}\,
	\frac{1}{n_d!}\, \left( \frac{\mu_d}{T} \right)^{n_d}\, 
\label{hichi}
\end{equation}

From this expansion, a series for baryonic susceptibility can be constructed.
Its radius of convergence gives the nearest critical point.  Successive
estimates for the radius of convergence can be obtained from these terms using
$ r_n = \sqrt{\big|\frac{\chi^n_B}{\chi^{n+2}_B}\big|}$ or $\rho_n = \big[\big|
\frac{\chi^0_B} {\chi^n_B}\big| \big]^\frac{1}{n}$.  We used terms up to 8th
order in $\mu$, i.e., estimates from 2/4, 4/6 and 6/8 terms of the series
eq. \ref{hichi}. 

The ratio $\chi_{11}/\chi_{20}$ can be shown \cite{endpt}  to yield the ratio
of widths of the measure in the imaginary and real directions at $\mu=0$.
This argument for the measure of the imaginary part of the fermionic 
determinant can be generalized to nonzero $\mu$ with some care,  by
constructing similarly as above the coefficients for the off-diagonal
susceptibility, $\chi_{11}(\mu)$. 

Before going on further to the results on the critical point, let us take a
detour of interest to the current heavy ion experiments.  Quark number
susceptibilities (QNS) which contribute the first nontrivial term in the above
Taylor expansion also have their own independent physical and theoretical
relevance.  They are crucial for some signatures  of quark-gluon plasma such as
fluctuations of charge and/or baryon number, and production of strangeness.
Their additional theoretical importance is due to the check they provide on
resummed perturbation expansions or any other scenario for the high $T$ phase.  

\subsection{ The Wr\'oblewski Parameter and Baryon-Strangeness Correlation}

\begin{figure}
\begin{minipage}{0.49\textwidth}
\includegraphics*[width=6.0cm]{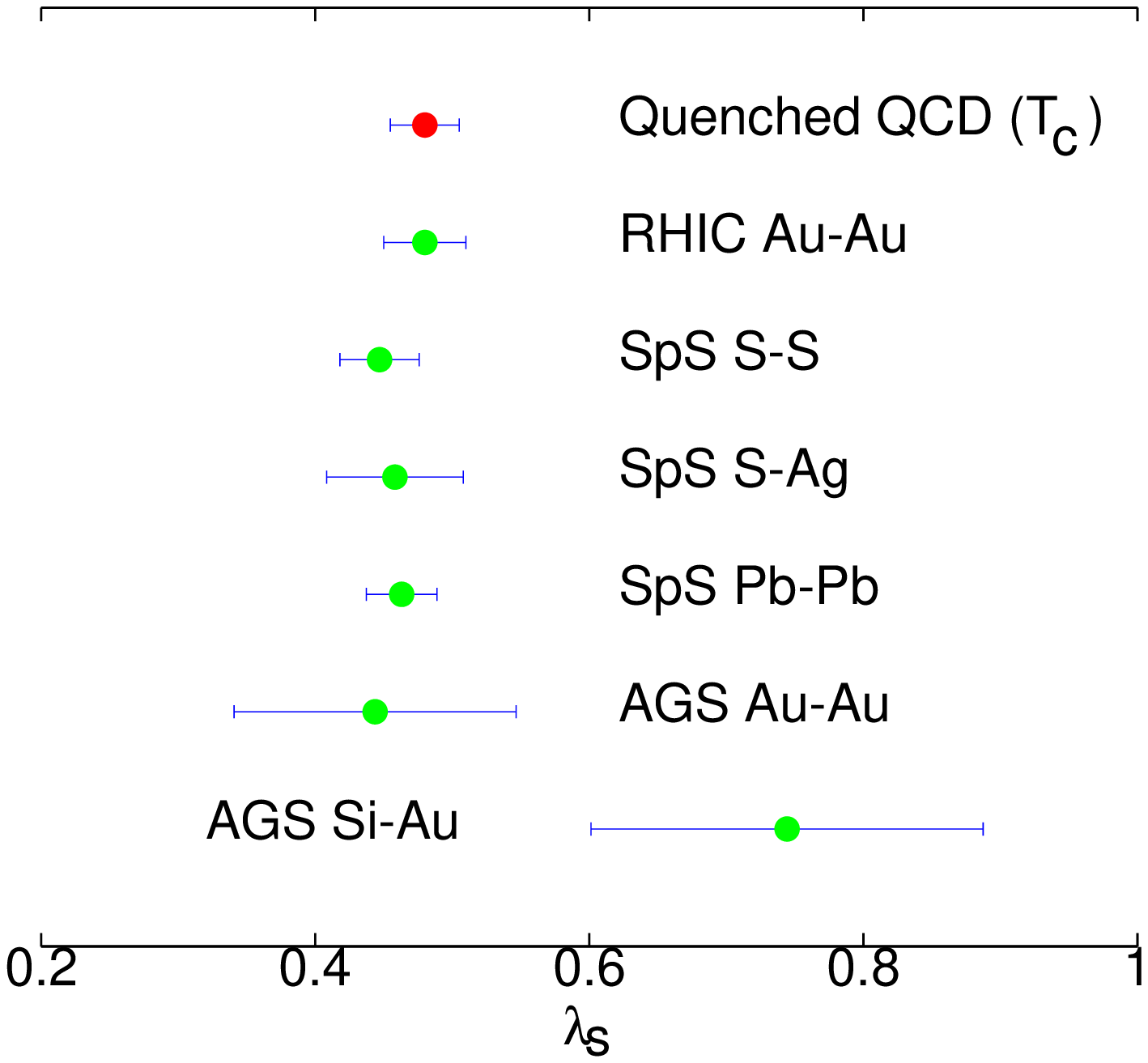}
\end{minipage}\hfill
\begin{minipage}{0.49\textwidth}
\includegraphics*[width=6.0cm]{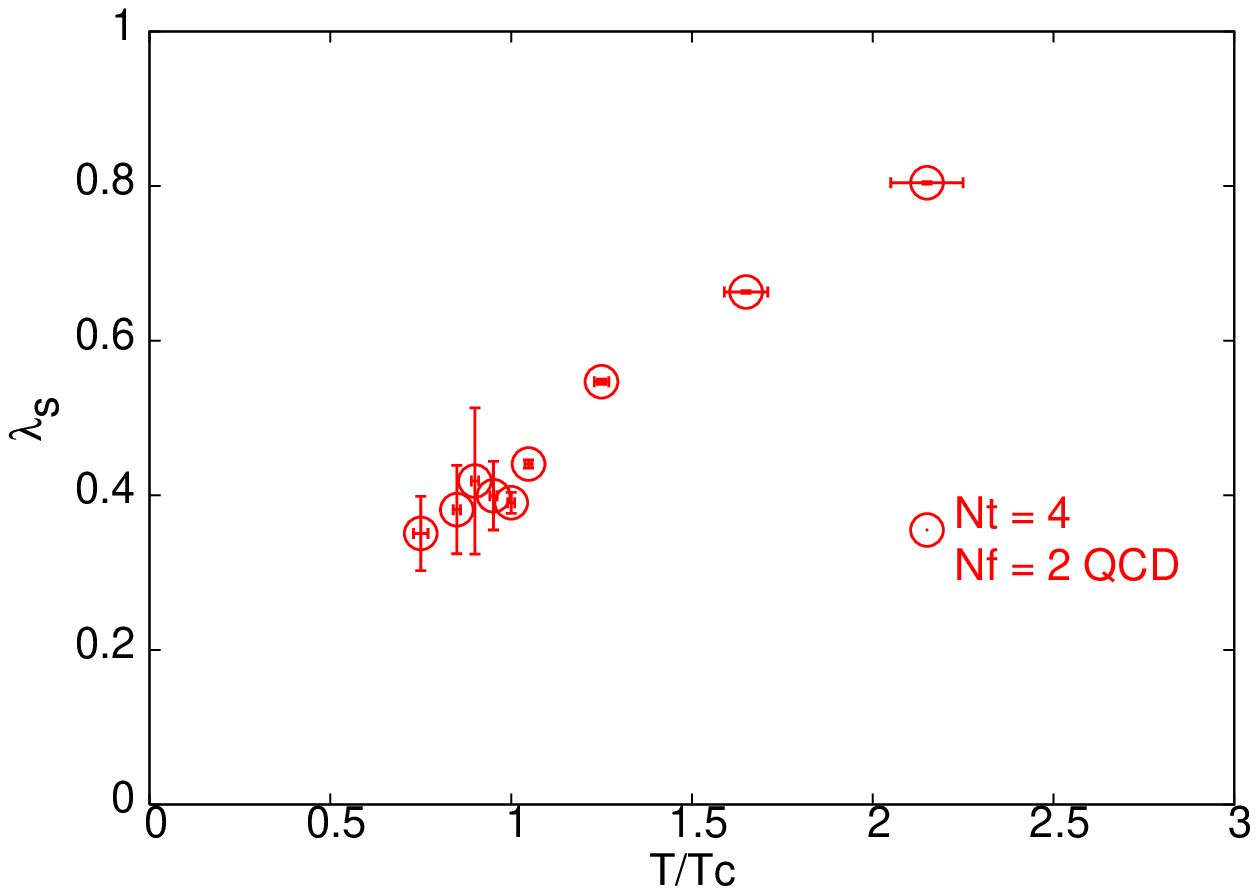}
\end{minipage} 
\caption{ Comparison of lattice results on the Wr\'oblewski Parameter
for quenched QCD with results from CERN and BNL (left). Corresponding full 
QCD results as a function of $T/T_c$ on the right. Taken from 
Refs. [4,24] respectively.}
\label{fg:five}
\end{figure}

Here we will touch upon two different aspects of the of physics hidden in the
QNS.  Ideally, one needs to obtain QNS in the continuum limit to  extract any
quantity of interest for heavy ion collisions.  It has recently been shown that
one can construct \cite{cbs} robust ratios which have negligibly small
theoretical and experimental systematic error.  The primary amongst them is the
Wr\'oblewski parameter.  Defined as  the ratio of the strange particles and the
non-strange particles produced in a collision, it has been studied widely as a
measure of the strangeness production.  Interestingly, most heavy ion collision
data seems to yield a factor two higher value for it than other hadronic
collisions.  Using the continuum values for QNS, and under certain assumptions
\cite{gg02}, one obtains $\lambda_s (T_c)  = 2\chi_s/(\chi_u +\chi_d) \approx
0.4-0.5$ which compares remarkably well with its latest RHIC value $0.47 \pm
0.4$, as shown in the left panel of Figure \ref{fg:five}. While these results
were obtained in quenched QCD in the continuum limit, also the full QCD
results \cite{cbs} shown in the right panel but on a small temporal lattice
($N_t =4$) are in very good agreement with these, as expected of a robust
ratio.

\begin{figure}[htb]
\begin{center}
\includegraphics*[width=7cm]{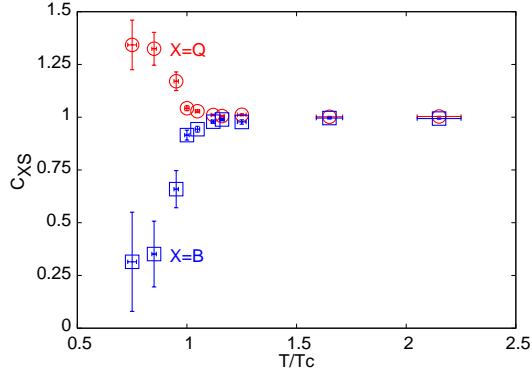}
\caption{$C_{BS}$ and $C_{QS}$ as functions of $T/T_c$.
   The quark masses used are $m_{ud}=0.1T_c$ and $m_s=T_c$.
   Taken from Ref. [24].}
\label{fg:six}
\end{center}
\end{figure}

QNS can also be put to use in studying the correlation between quantum numbers
$K$ and $L$ through the ratio $ C_{(KL)/L} = \frac{\langle KL\rangle-\langle
K\rangle\langle L\rangle}{ \langle L^2\rangle-\langle L\rangle^2} \equiv
\frac{\chi_{KL}}{\chi_L}$. Again being ratios, these too are expected to be
theoretically and experimentally robust. Such a strangeness-baryon number
correlation, $C_{BS} =  -3\,\frac{\chi_{BS}}{\chi_S} $, was proposed \cite{kmr}
as a distinguishing test between the various models/pictures of quark-gluon
plasma near $T_c$.  This, or the similarly defined strangeness-electric charge
correlation, $C_{QS}$, is expected to be unity if quarks are the sole carriers
of these quantum numbers (others being very heavy).  On the other hand, the
so-called sQGP model of Shuryak-Zahed \cite{SZ} predicts these to be 
0.66 and 1.2 respectively.  Our lattice results \cite{cbs} are exhibited in 
Figure \ref{fg:six}.

While one sees very different values for both $C_{BS}$ and $C_{QS}$ below
$T_c$, their rapid approach to unity clearly indicates the presence of
quark-like degrees of freedom immediately above $T_c$.  In particular, the
object which carries unit strangeness also has baryon number of -1/3 and a
charge of 1/3, just like a strange antiquark would. We have found that a
variation of the strange quark mass, $m_s/T_c$, between 0.1 and 1.0 does not
alter either the value for $T \ge T_c$, or its $T$-independence.  A natural
explanation of the $T$-behaviour arises if strange excitations with baryon
number become lighter at $T_c$. $T$-independence further suggests existence of
a single such excitation.  Such a picture is tantalizingly close to the
canonical expectations of a transition from hadrons to quarks.

\subsection{ The Critical Point}

\begin{figure}
\begin{minipage}{0.49\textwidth}
\includegraphics*[width=6.0cm]{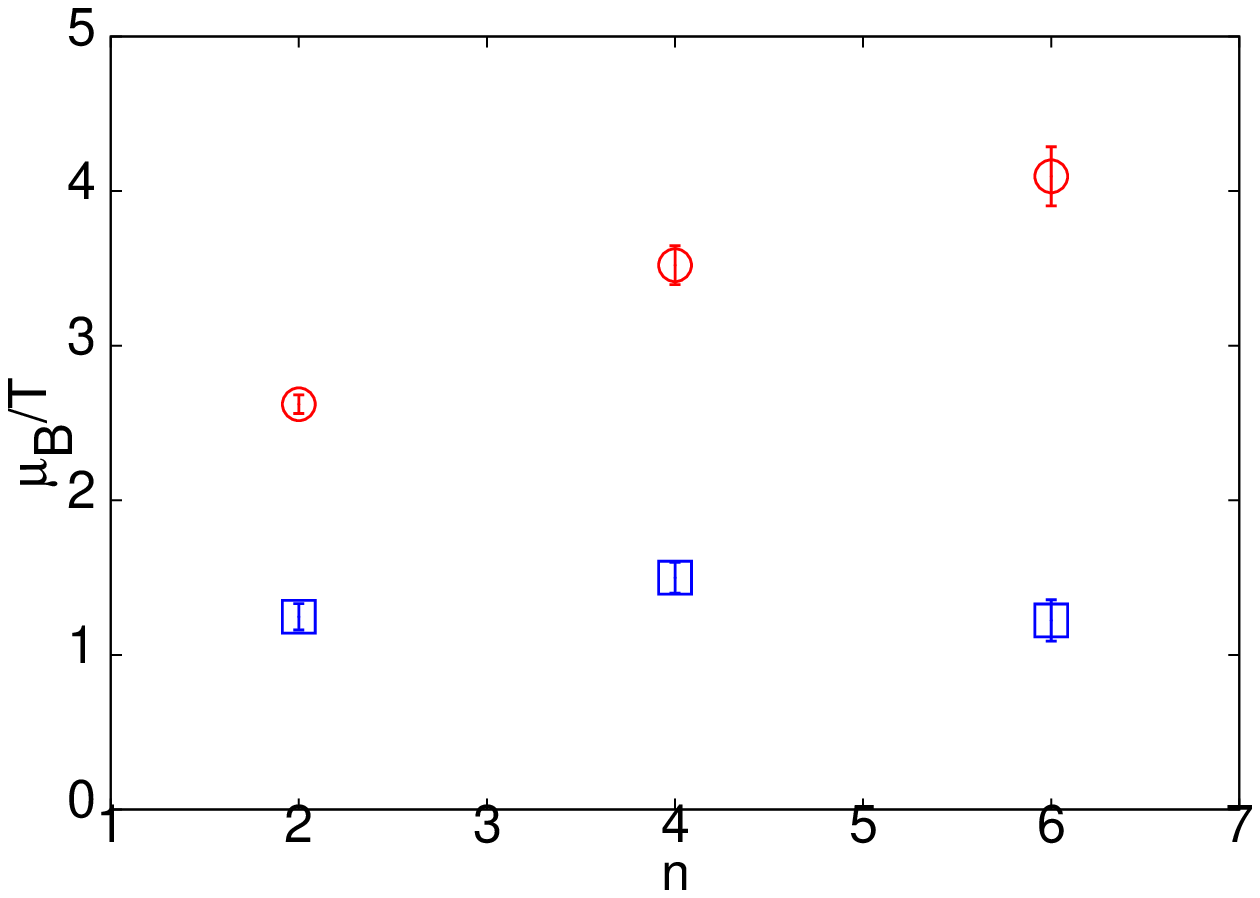}
\end{minipage}\hfill
\begin{minipage}{0.49\textwidth}
\includegraphics*[width=6.0cm]{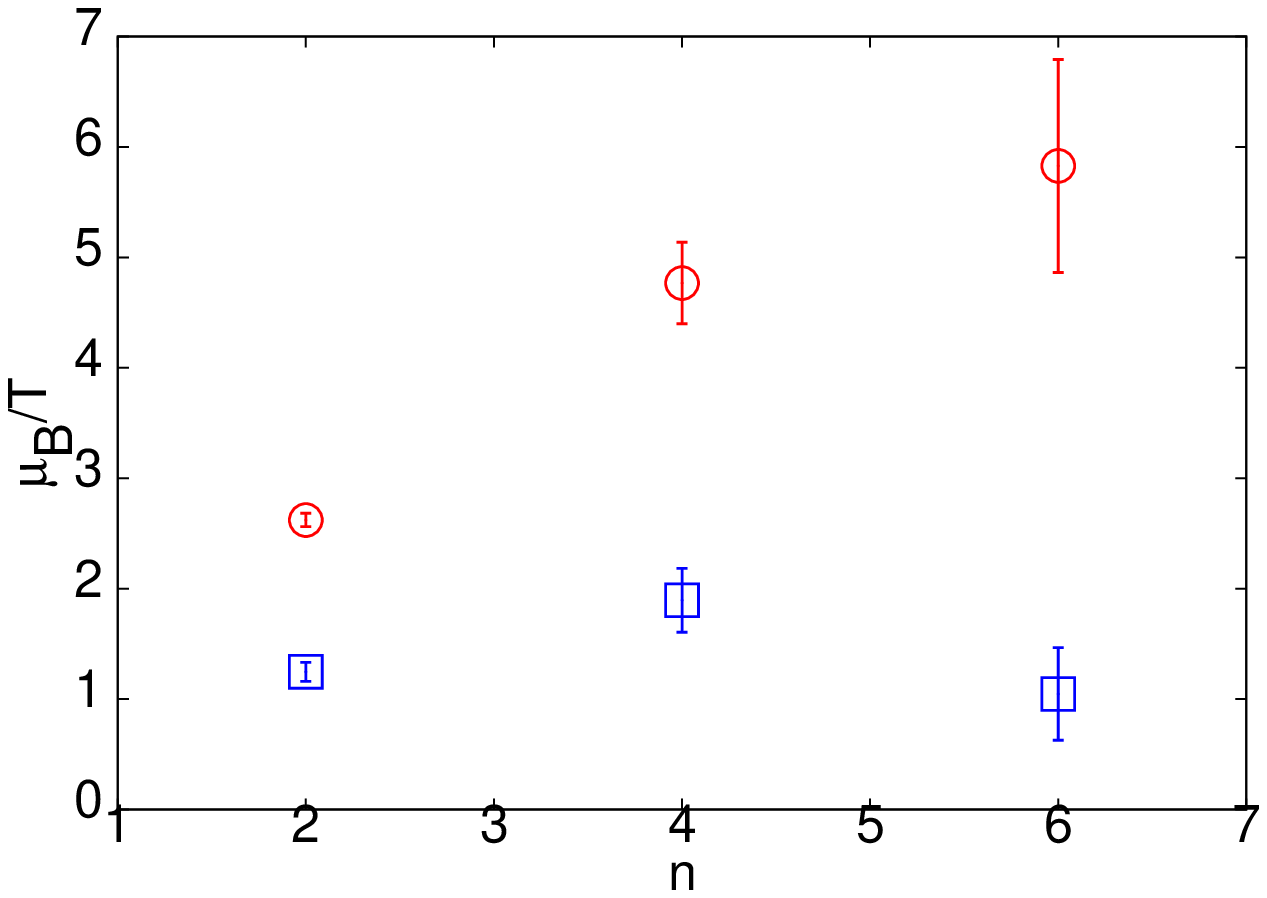}
\end{minipage} 
\caption{ Radii of convergence $\rho_n$ (left) and $r_n$ (right) as a function 
 of the order of the expansion at $T=0.95 T_c$ on a $4\times8^3$ lattice 
 (circles) and a $4\times24^3$ lattice (boxes).
   Taken from Ref. [21]. }
\label{fg:seven}
\end{figure}

Using both the definitions the radius of convergence above for the terms
up to the 8th order in $\mu$ in pressure, (6th order in baryon number
susceptibility), one obtains successively better estimates order by order.  This
can be done on each spatial volume, leading to a study of the volume dependence
of the radius of convergence.  Figure \ref{fg:seven} shows the results for 
both $\rho_n$ and $r_n$ on our smallest ($8^3$) and largest ($24^3$) lattice.
 
On the smaller lattice our results are consistent with the earlier estimate
using re-weighting method.  We observe strong finite size effects around $N_s
m_\pi \sim 6$ but our results on the largest lattice suggest a good stability
with respect to an increase in the order of the expansion. Extrapolation in $n$
leads to  the following estimate for the critical point : $\mu^E/T^E = 1.1 \pm
0.2$ at $T^E = 0.95T_c$.  Similar estimate has also been obtained for the
re-weighting method recently although the quark mass used ( and hence the mass
of the pion) was smaller in that case \cite{fk}.  Another attempt to look
\cite{BiC} for the critical point using terms only up to 6th order did not find
any critical point but they used much larger quark masses ( $m_\pi/m_\rho \sim
0.7$).

\begin{figure}[htb]
\begin{center}
\includegraphics*[width=8cm]{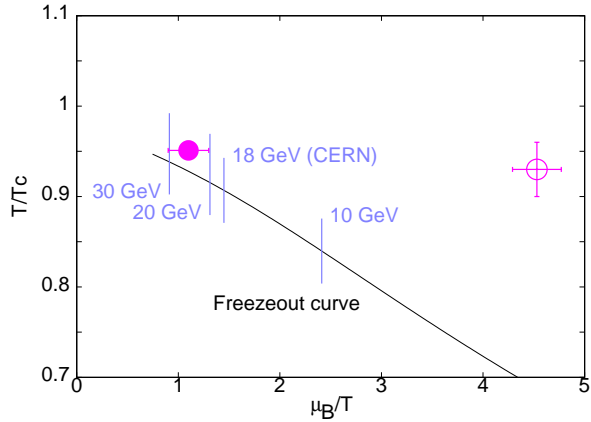}
\caption{QCD phase diagram and the freeze-out curve superimposed. The
   filled and open circles denote our [21] and earlier [19] 
   estimate of the critical point for nearly the same quark mass.  A scale of
   the CM energy per nucleon, $\sqrt S$, has been marked on the freeze-out
   curve. }

\label{fg:eight}
\end{center}
\end{figure}

\section{Summary}

Lattice QCD {\bf predicts} new states of strongly interacting matter and is
able to shed light on the properties of the Quark-Gluon Plasma phase.  One of
the major developments of the recent past in lattice QCD is the firming up of
the QCD phase diagram in $\mu$-$T$ plane on small $N_t$.  Different fermions
different methods of simulations for nonzero $\mu$, all lead to  good agreement
on the qualitative as well as the quantitative aspects.  All estimates of
$T_c$, and $(T_E, \mu_E)$ are mutually consistent when compared for the right
quark masses and in the thermodynamic limit.  Our estimate for the critical
point  is $\mu_B/T \sim 1-2$, as shown in Figure \ref{fg:eight}.  Also shown in
the figure is a freeze-out curve \cite{pvt}, converted to a value of $T_c$
appropriate to our computation.  Such a curve results from the analysis of the
heavy-ion data on particle yields.   The required collision energy to reach the
appropriate point are marked on the figure, which indicate the exciting
possibility of discovering the critical point in a low energy RHIC run in
the near future. 

Various physical quantities have been obtained in the continuum limit in the
quenched approximation to QCD.  These include the equation of state, the
specific heat, the speed of sound in the neighbourhood of $T_c$ and the quark
number susceptibilities.  While the former are needed in hydrodynamical
analysis of the particle spectra, and the resultant collective flow, the latter
(QNS) are directly relevant to the physics of quark-gluon plasma signals at
RHIC.  The quenched lattice QCD estimate of Wr\'oblewski parameter,
$\lambda_s$, which is a measure of strangeness production in heavy ion
collision experiments, has been known to be in excellent agreement with the
RHIC and SPS results. 
 
Strong efforts are going on to extend these results to full QCD. Many
features seem to change very little, although the major change in form of the
order of the phase transition leads to quantitative differences near $T_c$.
Interestingly the robust Wr\'oblewski parameter seems also not to change
quantitatively near $T_c$, although one still needs more precise results
and that too for realistic pion and kaon masses. Our results on 
baryon number-strangeness and electric charge-strangeness correlations suggest
i) a rapid change in going through the transition in full ($N_f=2$) QCD, and
ii) the quark-gluon plasma to have quark-like excitations even close to $T_c$.

The heavy ion data from CERN and BNL has provided us a lot of surprises, and
will continue to do so in future.  Lattice QCD has played an important part
in understanding some of these, and clearly a lot more work is ahead both
in form of better precision as well as new ideas for experimentally measurable
lattice predictions.

\end{document}